\documentclass[aps,prb,reprint,superscriptaddress]{revtex4-2}

\usepackage{amsmath}
\usepackage{amssymb}
\usepackage{xcolor}

\usepackage{amsmath,amssymb,amsfonts,bm,color,graphicx}

\usepackage{tabularx,multirow,array,diagbox}

\usepackage{bm}

\usepackage[unicode=true,colorlinks=true]{hyperref}

\usepackage[etex=true,export]{adjustbox}

\usepackage{ulem}
\newcommand{\beq}{\begin{equation}}
\newcommand{\eeq}{\end{equation}}
\newcommand{\beqn}{\begin{eqnarray}}
\newcommand{\eeqn}{\end{eqnarray}}

\begin{document}
\title{Parity-protected superconducting qubit based on topological insulators}

\author{Guo-Liang Guo}
	\affiliation{School of Physics and Institute for Quantum Science and Engineering, Huazhong University of Science and Technology, Wuhan, Hubei 430074, China}
	\affiliation{Wuhan National High Magnetic Field Center and Hubei Key Laboratory of Gravitation and Quantum Physics, Wuhan, Hubei 430074, China}
 
\author{Han-Bing Leng}
\affiliation{Key Laboratory of Functional Materials and Devices for Informatics of Anhui Higher Education Institutes, Fuyang Normal University, Fuyang, 236037, China}

	\author{Xin Liu}
	\email{phyliuxin@hust.edu.cn}
	\affiliation{School of Physics and Institute for Quantum Science and Engineering, Huazhong University of Science and Technology, Wuhan, Hubei 430074, China}
	\affiliation{Wuhan National High Magnetic Field Center and Hubei Key Laboratory of Gravitation and Quantum Physics, Wuhan, Hubei 430074, China}

\begin{abstract}
We propose a novel architecture that utilizes two 0-$\pi$ qubits based on topological Josephson junctions to implement a parity-protected superconducting qubit. The topological Josephson junctions provides protection against fabrication variations, which ensures the identical Josephson junctions required to implement the0-$\pi$ qubit. By viewing the even and odd parity ground states of a 0-$\pi$ qubit as spin-$\frac{1}{2}$ states, we construct the logic qubit states using the total parity odd subspace of two 0-$\pi$ qubits. This parity-protected qubit exhibits robustness against charge noise, similar to a singlet-triplet qubit's immunity to global magnetic field fluctuations. Meanwhile, the flux noise cannot directly couple two states with the same total parity and therefore is greatly suppressed. Benefiting from the simultaneous protection from both charge and flux noise, we demonstrate a dramatic enhancement of both $T_1$  and $T_2$ coherence times. Our work presents a new approach to engineer symmetry-protected superconducting qubits.
\end{abstract}

\maketitle
\section{introduction}
Superconducting circuits based on Josephson junctions are a leading platform for quantum computation \cite{Makhlin2001,Bouchiat2003,You2005,Arute2019,Tsioutsios2020,Dborin2022,Wu2021,Gyenis2021,Rymarz2021,Siddiqi2021,Acharya2023}. In recent decades, significant progress has been made in coherence times\cite{Gyenis2021a,Lisenfeld2023}, and in demonstrating high-fidelity single- and two-qubit gates \cite{Paik2011,Barends2013,Casparis2016,Nesterov2018,Place2021,Bao2022,Chang2023,Chen2023}. These advances are enabled not only by technological improvements but also by innovations in qubit designs \cite{Gladchenko2009,Smith2020,Kalashnikov2020,Schrade2022}. A milestone in superconducting qubit development was the introduction of the Transmon qubit \cite{Koch2007}, which provides remarkable protection against dephasing. More recently, the qubit architecture has been developed based on the concept of symmetry protection, such as 0-$\pi$ qubit \cite{Kitaev2006,Brooks2013,Dempster2014,Groszkowski2018,Weiss2019,Paolo2019}, fluxonium qubit\cite{Manucharyan2009,Pop2014,Nguyen2019}, which aim to realize hardware with intrinsic robustness against noise. Taking the parity-protected qubit\cite{Larsen2020,Guo2022} for example, its qubit states are protected by Cooper pairs parity, suppressing charge noise due to vanishing transition matrix elements, protecting the qubit states from depolarizing. However, the current experimental setup remains sensitive to flux noise which can weaken the parity protection. Therefore, constructing a qubit architecture that can suppress both charge and flux noise is essential for advancing superconductor-based quantum computation. Moreover, progress in materials science provides additional degrees of freedom, such as spin\cite{Szombati2016,Yamashita2019,Fukaya2022,Mazanik2022,Ness2022} and edge states\cite{Fu2008,Fu2009,Heck2011,Wiedenmann2016,Schmitt2022a,Schmitt2022,OldeOlthof2023,Fu2023}, for controlling Josephson junctions. These capabilities create opportunities to improve qubit designs and enable symmetry or topology protection of quantum computation at the hardware level.

\begin{figure}[!htbp]
	\centering
	\includegraphics[width=1\columnwidth]{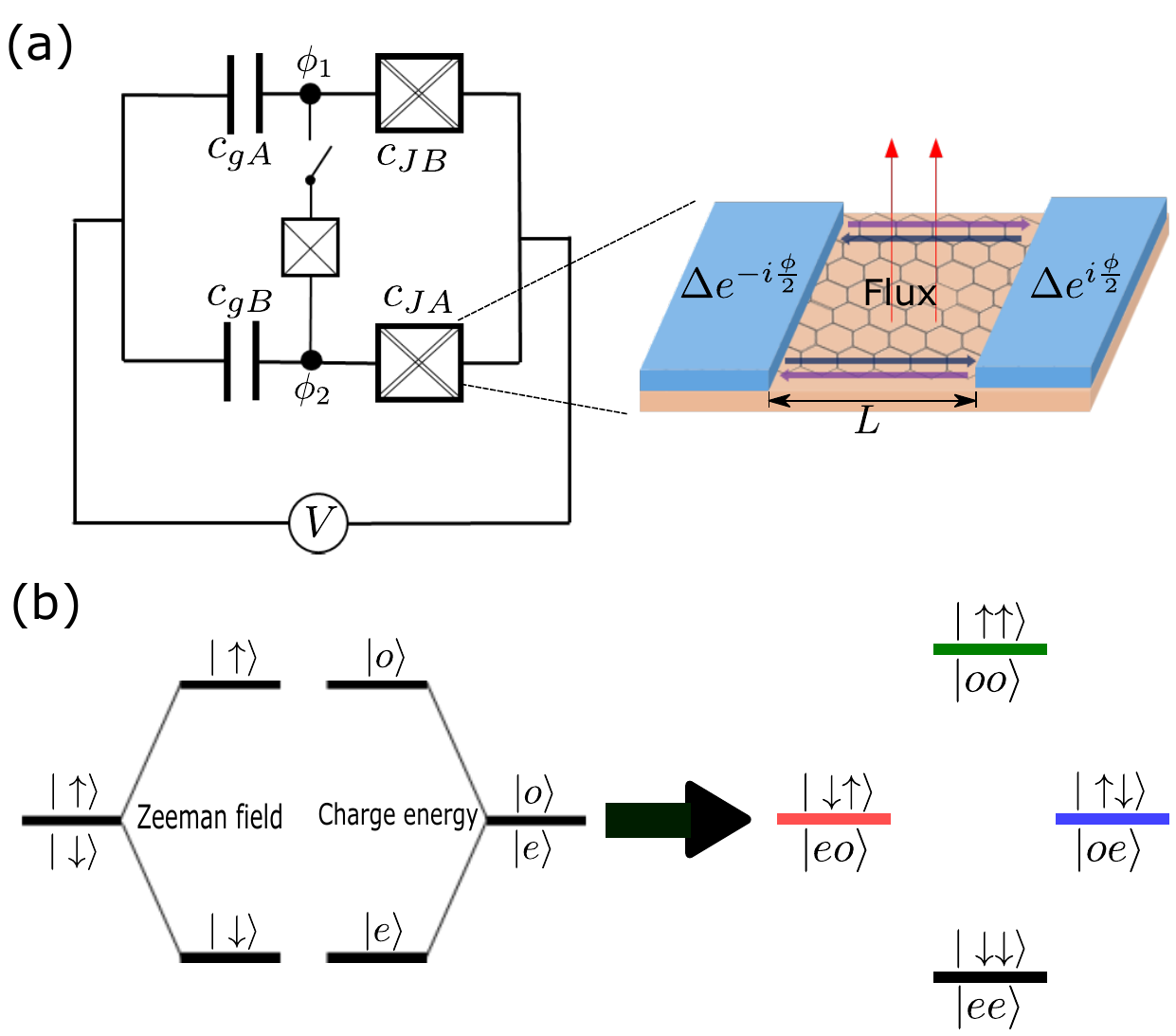}
	\caption{(a) Circuit for the parity-spin qubit, the double cross represents pairs of cooper pairs tunneling, the normal Josephson Junction with switch can be gain by gatemon. The right part is the schematic diagram of TI Josephson Junction with flux in Junction area. (b) Analogy of single $0-\pi$ qubit and single spin-1/2 parity.}
\label{fig-1}
\end{figure}

In this work, we propose a novel parity-protected qubit architecture utilizing two coupled 0-$\pi$ qubits, as shown in Fig.~\ref{fig-1}(a). Each 0-$\pi$ qubit is based on a topological Josephson junction made from a two-dimensional topological insulator (2DTI). These junctions exhibit nearly degenerate parity-protected states ${|\rm{e}\rangle,|\rm{o}\rangle}$ which can be viewed as pseudo-spin-$\frac{1}{2}$ systems, analogous to spin up and down (Fig.~\ref{fig-1}(b)). Notably, the time-reversal symmetry protected helical edge states are robust against the fabrication variations such as disorders or junction geometries and greatly enhance the identity of the 0-$\pi$ qubits. Bringing two 0-$\pi$ qubits together allows us to construct the logic qubit states $|\rm{eo}\rangle$ and $|\rm{oe}\rangle$ from the total parity odd subspace, similar to a conventional singlet-triplet spin qubit (Fig~\ref{fig-1}(b)). The energy splitting between $|\rm{e}\rangle$ and $|\rm{o}\rangle$ can be tuned by the offset charge $n_g$ and the ratio of charging energy $E_c$ to Josephson coupling $E_J$ (Fig.~\ref{fig-1}(b)). Our analysis verifies that parity-protected qubit architecture can simultaneously suppress charge and flux noise while maintaining sufficient anharmonicity. The estimated $T_1$ and $T_2$ coherence times can reach hundred milliseconds. In addition, we propose the operation scheme following the operation of S-T qubit, achieving (partial) electrical control of the qubit states.

\section{Parity protected Superconducting qubit with TI Josephson Junction}
\begin{figure*}[!htbp]
	\centering
	\includegraphics[width=17cm]{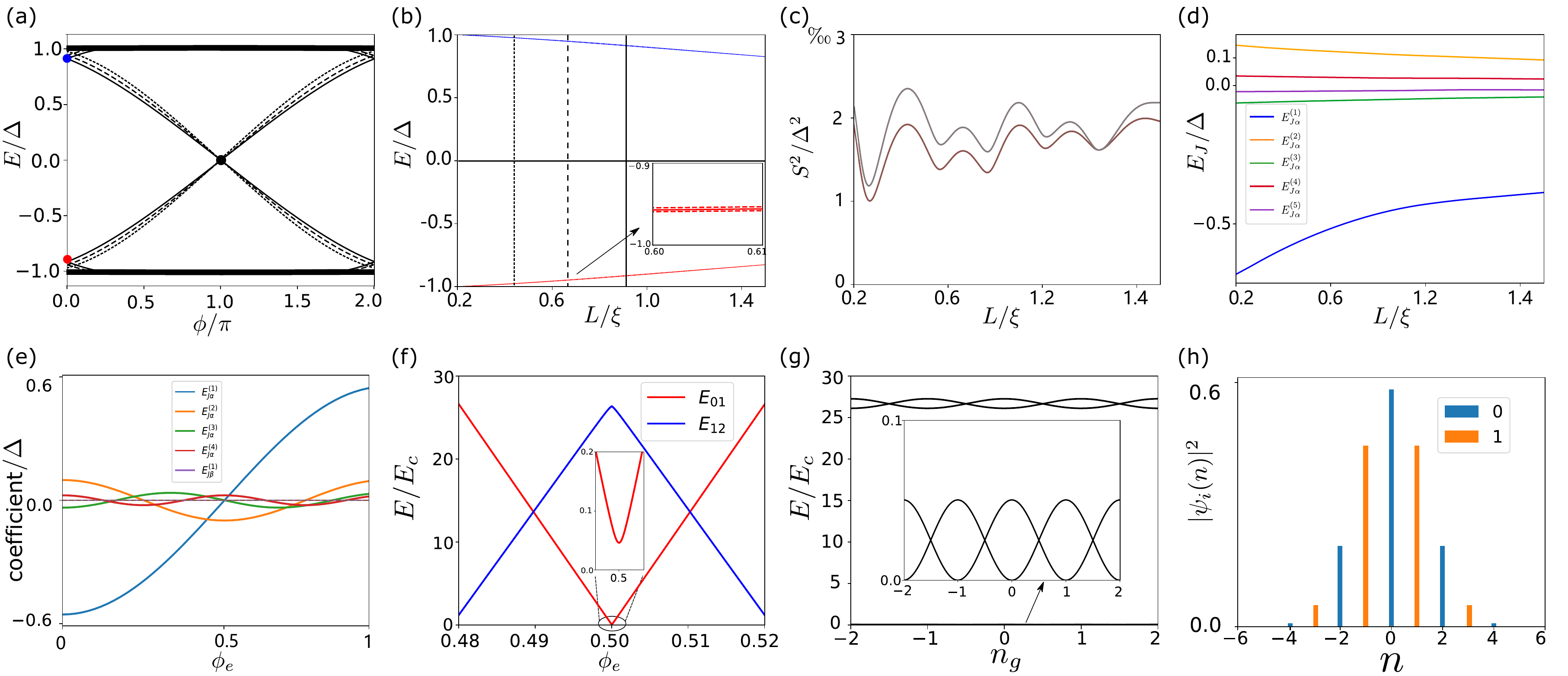}
	\caption{(a) The Andreev bound states(ABS) of TI Josephson junction for different Junction lengths, corresponding to the vertical line in (b). (b) The energy changes with the junction length, the three solid lines correspond to the three points in (a), and the dashed lines are the same case in the presence of disorder. (c) The variance of the eigenenergies at $\phi=0$ changes with junction length in the presence of disorder. (d) The Fourier coefficient of Josephson potential changes with Junction length. (e) The Fourier coefficient of the Josephson potential changes with flux($\phi_e$) in the short junction limit($L\approx0.65\xi$). (f) The energy difference($E_{ij}=E_j-E_i$) changes with flux. (g) The lowest four state energies change with offset charge($n_g$) at half flux quantum. (h) The lowest two states distribution in charge space. (f)(g)(h) are gain with data in (e) with $E_{J}^{(2)}/E_c\approx 40$.}
\label{fig-2}
\end{figure*}

The TI Josephson Junction is composed of a 2D TI that is covered with superconductors on both ends, as depicted in Fig.~\ref{fig-1}(a). Here, We take the BHZ model as an example\cite{Koenig2008,Qi2010} of the topological insulator. In basis $(|E_1+\rangle,|H_1+\rangle,|E_1-\rangle,|H_1-\rangle)^T$, the Hamiltonian takes the form 
\begin{equation}
\begin{aligned}
\mathcal{H} &=\left(\begin{array}{cc}
h(k) & 0 \\
0 & h^{*}(-k)
\end{array}\right) \\
\end{aligned}
\end{equation}
where the two blocks of $\mathcal{H}$ represent the spin up and down components, respectively, and $h(k) =\epsilon(k) \mathrm{I}_{2 \times 2}+d_{a}(k) \sigma^{a}$, $\epsilon(k) =C-2D\left(2-\cos(k_x)-\cos(k_y)\right)$, $d_{a}(k)=\left(A \sin(k_x),-A\sin(k_y), M(k)\right)$, $M(k)=M-2B\left(2-\cos(k_x)-\cos(k_y)\right)$, $\sigma$ acting on orbital space. In the topological phase($-4B<M<4B$), there exist helical edge states. In the 2DTI Josephson junction, the Fermi surface can be tuned by a gate voltage and when it lies in the bulk energy gap, only the helical edge states can support Cooper pairs tunneling. As the junction respect time-reverse symmetry, each edge state can support Andreev Bound states (ABS) independently, leading to the ABS with a large Josephson coupling coefficient in the short junction limit. The system maintains time-reverse symmetry and particle-hole symmetry when the superconducting phase difference is $\pi$(Fig.~\ref{fig-2}(a) black dot), the eigenenergies of the system are exactly zero and are independent of the junction length(black line in Fig.~\ref{fig-2}(b)). In this case, the Andreev level in short junction limit can be described with the equation\cite{Beenakker1991,Fu2008}
\begin{equation}
    E(\phi)=\tilde{\Delta}\cos(\frac{\phi}{2}),
\end{equation}
where $\tilde{\Delta}$ relies on the junction length and it converges to the superconductor gap ($\Delta$) in the short junction limit. The shape of Andreev level is mainly determined by the eigenenergies at $\phi=0$ (Fig.~\ref{fig-2}(a) red dot and blue dot). In Fig.~\ref{fig-2}(b)(the solid red line and blue line), we plot the eigenenergies at $\phi=0,\pi$, which reflects the Josephson potential, changes with junction length. It clearly shows that the eigenenergies are slightly rely on the junction length, which indicate the Josephson potential shape can immune to geometry difference in 2DTI Josephson junctions. Meanwhile, there may exist disorder effect in experiment, we found that the effect is significantly small and can be neglected, as shown in the dashed lines in Fig.~\ref{fig-2}(b). It presents the eigenergies change with Junction length in the presence of disorder (random onsite potential $V_{\rm{dis}}s_0\sigma_0$) with disorder strength ranges from $V_{\rm{dis}}=-5\Delta$ to $V_{\rm{dis}}=-5\Delta$, $\Delta$ is the superconducting gap. Though the disorder effect may change the equivalence between the two edge states of 2DTI, it is significantly small. And in Fig.~\ref{fig-2}(c), we plot the standard deviation ($S^2$) of the eigenenergies at $\phi=0$ as a function of junction length in the presence of disorder, which indicates the little changes caused by the disorder, it means that 2DTI Josephson junction can immune to disorder effect. Moreover, as the disorder does not break the time-reversal symmetry, the Josephson potential remains an even function in $\phi$, keeping zero $\sin(\hat{\phi})$ term of the Josephson potential. The corresponding Fourier components of the Josephson potential are shown in Fig.~\ref{fig-2}(d), exhibiting a polynomial rather than exponential dependence on the Junction length. In our calculation, we take the BHZ model and Al superconductor as an example, the coherent length is about $\xi=\hbar v_f/2\Delta\approx1000$ nm\cite{Schaepers2001}. With state-of-the-art technology, the length difference of two TI Josephson Junctions can be smaller than $1\%\xi$. Consequently, the Josephson coupling difference is significantly small as shown in the inset of Fig.~\ref{fig-2}(b). And this conclusion holds for other quantum spin hall materials\cite{Yang2022,Shumiya2022}. Therefore, It is feasible to obtain multiple nearly identical Josephson Junctions with nearly the same Josephson coupling energies. The conclusion can be applied to all 2DTI Josephson Junction systems and independent of the 2DTI materials. It is beneficial to qubit design, such as Fluxonium\cite{Manucharyan2009,Pop2014,Krantz2019} and parity-protected qubit\cite{Smith2020,Larsen2020,Kalashnikov2020} which commonly needs two or more identical Josephson Junctions.



With 2DTI Josephson junction, we can construct 0-$\pi$ qubit in one Josephson junction. In the 2DTI Josephson Junction, the two edge states can support two paths for Cooper pairs tunneling, if there exists flux in the junction area $\phi_{\rm{ext}}=2\pi\phi_e/\phi_0$, $\phi_0=h/2e$, the two paths for Cooper pairs tunneling will become interference. At half flux quantum($\phi_e=\frac{1}{2}\phi_0$), the odd number of Cooper pairs tunneling across the junction will be coherently destructive, leaving only the even number of Cooper pairs tunneling. At half flux quantum, the general form of the Josephson potential takes the form\cite{Larsen2020,Guo2022}
\begin{equation}
V_{J}(\hat{\phi})=\sum_{m} E_{J}^{(m)} \cos (m \hat{\phi}),
\end{equation}
where $\hat{\phi}$ is the superconducting phase difference. At half flux quantum, the potential is only left with $\cos(m\hat{\phi})$ with even values of m. With the Josephson potential, we can write the qubit Hamiltonian
\begin{equation}
H=\sum_{n n' m}\left(4 E_{c}\left(\hat{n}-n_{g}\right)^{2} \delta_{n n'}+\frac{E_{J}^{(m)}}{2} \delta_{n, n'+m}\right)|n\rangle\langle n'|+h.c,
\end{equation}
where $E_c$ is the charge energy, $\hat{n}$ is the Cooper pair number, $n_g$ is the offset charge, and $E_{J}^{m}/2$ corresponds to the $m$-th nearest hopping due to the $m$ Cooper pairs tunneling simultaneously. Several leading terms of the potential are plotted in Fig.~\ref{fig-2}(a). At half flux quantum, there only exists even number of Cooper pair tunneling, corresponding to $E_{J}^{(2)}\cos2\hat{\phi}$ and higher terms, corresponding to $\pi$ period Josephson potential. Due to the Time-reversal symmetry of the two edge states, there is little $\sin m\hat{\phi}$ term in the potential. And we get two identical potential wells at $\phi=0,\pi$, supporting two nearly degenerate states. The eigenenergy difference $E_{ij}=E_j-E_i$ changes with flux is shown in Fig.~\ref{fig-2}(f). At half flux quantum, there exist two nearly degenerate states with a large energy gap(about$\sqrt{32E_{J}^{(2)}E_c}$) separating them from the higher energy levels. Moreover, the lowest two states are not exactly degenerate but have a finite gap $\delta E$ due to the charge energy. The gap is approximate with the form\cite{Smith2020} 
\begin{equation}
\delta E \approx 16 E_{c} \sqrt{\frac{2}{\pi}}\left(\frac{2 E_{J }^{(2)}}{E_{c}}\right)^{3 / 4} e^{-\sqrt{2 E_{J}^{(2)} / E_{c}}}.
\end{equation}
The several lowest energy levels change with offset charge($n_g$) are shown In Fig.~\ref{fig-2}(g). To characterize the wavefunction of the single 0-$\pi$ qubit states, in Fig.~\ref{fig-2}(h), we plot the wavefunction distribution of the lowest two states in charge space, it clearly shows that the wavefunction distribution is localized either on even number sites or odd number sites, labeled as $|\rm{e}\rangle,|\rm{o}\rangle$. This is because potential well at $\phi=0,\pi$ are identical, the hopping between opposite parity is forbidden which can protect 0-$\pi$ qubit states. Moreover, the consistency of the 2DTI Josephson Junction is topologically protected, it can protect the states from imperfect fabrications. 

Then, we consider the qubit properties. The charge noise and flux noise are the main noise source in 2DTI Josephson Junction. The charge noise can be suppressed by increasing $E_{J}^{(2)}/E_c$ similar to transmon\cite{Koch2007}, but it will increase the sensibility of flux\cite{Guo2022}. At half flux quantum, the qubit states are very sensitive to flux noise (Fig.~\ref{fig-2}(f)). As the deviations from half flux quantum($\delta\phi_e$) will introduce the term $E_J^{(1)}(\delta\phi_e)\cos(\hat{\phi})$, it can couple the qubit states($E_J^{(1)}(\delta\phi_e)\langle 0|\cos(\hat{\phi})|1\rangle$) directly. Additionally, the first-order derivative of $E_J^{(1)}(\phi_e)$ reaches its maximum at half flux quantum, which means large $E_J^{(1)}(\delta\phi_e)$, further deteriorating the qubit states. 
So, the protection needs to be improved further.

\section{Parity spin qubit}

\subsection{Model Hamiltonian}

As single 0-$\pi$ qubit has two lowest states $|\rm{e}\rangle$ and $|\rm{o}\rangle$, which can be treated as single spin-$\frac{1}{2}$ particle, we can enhance the qubit properties with two sets of 0-$\pi$ qubits, similar to S-T qubit. We then focus on the parity-spin qubit realized with 2DTI Josephson Junctions, which can improve charge noise and flux noise simultaneously.
\begin{figure}[!htbp]
	\centering
	\includegraphics[width=1\columnwidth]{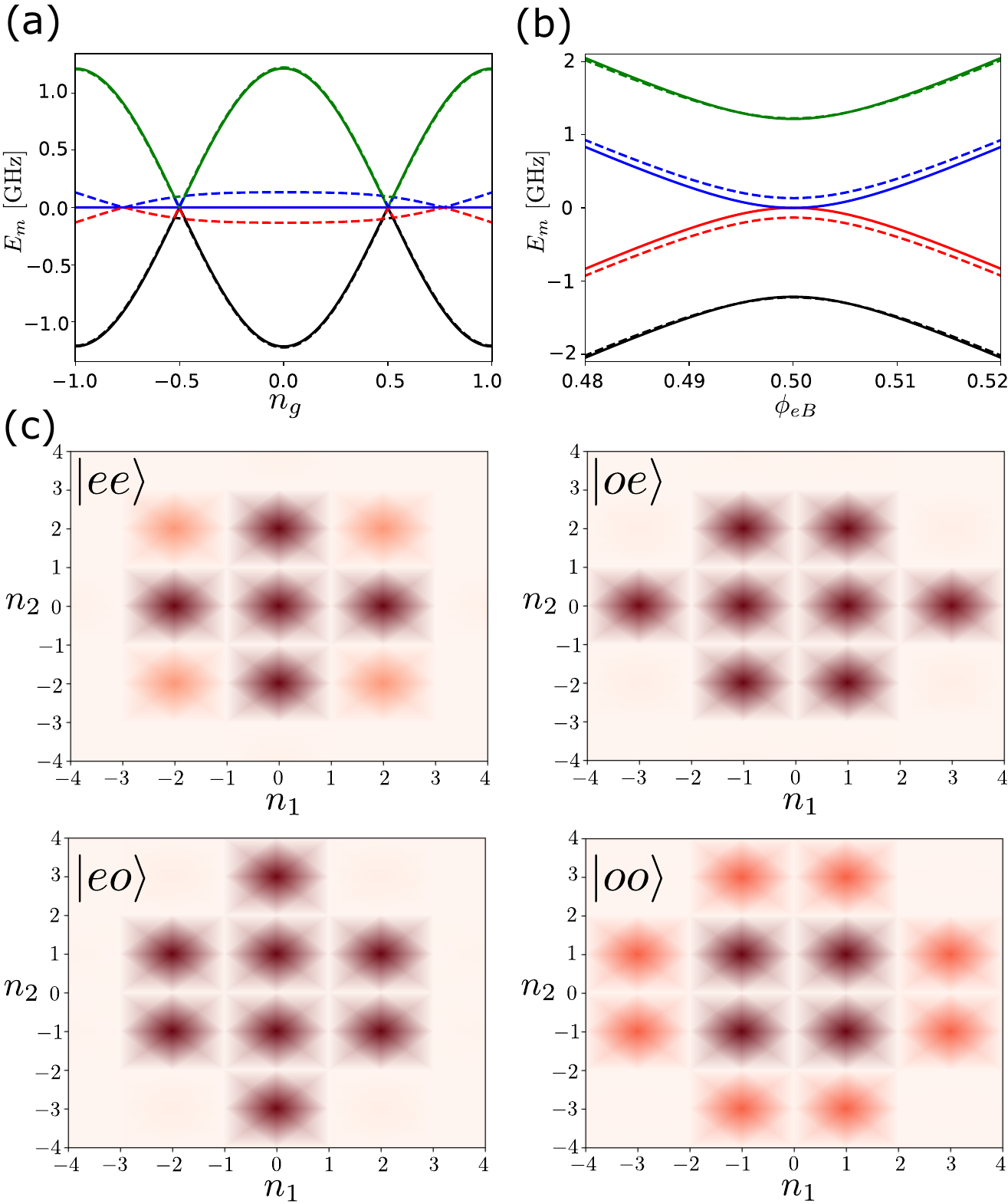}
	\caption{(a)(b) The two TI JJ have the same parameters with $E_{J}^{(2)}/E_c=9$. (a) The lowest four states energies change with $n_g$ at the condition $\phi_{eA(B)}=0.5$. (b) The four states' energy changes with $\phi_{eA(B)}$ at the condition $n_g=0$ and $\phi_{eB(A)}=0.5$, dashed is the same case with $x=\delta C_g /C_g=0.5$. (c) The wavefunction of the four states.}
\label{fig-3}
\end{figure}
Based on the consistency of the 0-$\pi$ qubit with 2DTI Josephson Junction, we consider the circuit configuration as shown in Fig.~\ref{fig-1}(a), two 0-$\pi$ qubits are shunted by a capacitor. The voltage is applied to tune the two offset charges, which can change the energy difference of the two 0-$\pi$ qubits states, simultaneously, analogous to a uniform magnetic field in the S-T qubit. The two nodes are connected by a normal semiconductor Josephson Junction with a switch which can be controlled using a gatemon in the experiment setup. In our proposal, the gatemon is used to manipulate the qubit states. We assume the switch is initially in the off state and define the flux node variables as ${\phi_1,\phi_2}$. Then, the circuit Lagrangian takes the form
\begin{equation}
\begin{aligned}
    \mathcal{L}&=\frac{1}{2}C_{JA}\dot{\phi_1}^2+\frac{1}{2}C_{JB}\dot{\phi_2}^2+\frac{1}{2}C_{gA}(V-\dot{\phi}_1)^2+\frac{1}{2}C_{gB}(V-\dot{\phi}_2)^2\\
    &+E_{JA}^{(1)}(\phi_{eA})\cos(\hat{\phi}_1)+E_{JA}^{(2)}(\phi_{eA})\cos(2\hat{\phi}_1)\\
    &+E_{JB}^{(1)}(\phi_{eB})\cos(\hat{\phi}_2)+E_{JB}^{(2)}(\phi_{eB})\cos(2\hat{\phi}_2),
\label{eq1}
\end{aligned}
\end{equation}
where $C_{JA(B)}$ are the junction capacitance, $C_{gA(B)}$ are the capacitance of the two 
shunted capacitors. The first line is the kinetic energy contributed by all the capacitors, and the other term is the Josephson coupling energy of the two junctions, which can be independently tuned by the flux $\phi_{eA(B)}=2\pi\Phi_{A(B)}/\Phi_0$ in the Junction area, respectively. We assume that the two junctions are identical(the consistency of 2DTI Josephson junctions) which means $E_{JA}^{(i)}(\phi_e)=E_{JB}^{(i)}(\phi_e)$ and we label it as $E_{J}^{(i)}$, $i$ denotes the order of the cosine term. Without loss of generality, we consider the Josephson potential only contains $\cos\phi$ and $\cos(2\phi)$ terms. Fig.~\ref{fig-2}(e) shows the coefficient changes with flux. In experiment, the superconductor is usually made of aluminum, whose superconducting gap ($\Delta$) is about 44 GHz\cite{Larsen2020}, then the coefficient can be expressed with the expression
\begin{equation}
\begin{aligned}
    E_{J}^{(1)}(\phi_e)&=-25\cos(\pi\phi_e)\\
    E_{J}^{(2)}(\phi_e)&=4.5\cos(2\pi\phi_e).
\end{aligned}
\label{eq2}
\end{equation}
At the 0-$\pi$ point ($\phi_e=0.5$), the two junctions can only allow even number Cooper pair tunneling, corresponding $E_{J}^{(1)}(0.5)=0$ in Eq.~(\ref{eq2}), only left with $E_{J}^{(2)}\cos(2\phi)$ term. With the Lagrangian Eq.~(\ref{eq1}), we can gain the Hamiltonian by doing the Legendre transformation, it takes the form 
\begin{equation}
\begin{aligned}
    H=&4E_{cA}(\hat{n}_{1}-n_{gA})^2+4E_{cB}(\hat{n}_{2}-n_{gB})^2\\
    &+E_{J}^{(1)}(\phi_{eA})\cos(\hat{\phi_1})+E_{J}^{(1)}(\phi_{eB})\cos(\hat{\phi_2})\\&+E_{J}^{(2)}(\phi_{eA})\cos(2\hat{\phi_1})+E_{J}^{(2)}(\phi_{eB})\cos(2\hat{\phi_2}),
\label{eq3}
\end{aligned}
\end{equation}
where $E_{cA(B)}=e^2/2(C_{JA(B)}+C_{gA(B)}), [\hat{\phi}_j,\hat{n}_j]=i(j=1,2)$, $n_{gA(B)}=C_{gA(B)}V/2e$ is tuned by the same voltage. The Hamiltonian is the sum of two 0-$\pi$ qubit Hamiltonian, each 0-$\pi$ qubit Hamiltonian can support two parity protected states$\{|\rm{e}\rangle,|\rm{o}\rangle\}_{A(B)}$ in the transmon regime or weak transmon regime($E_{J}^{(2)}/E_c>1$). The two states has large energy gap with the higher energy level, as shown in Fig.~\ref{fig-2}(f)(h). In low energy basis $\{|\rm{e}\rangle,|\rm{o}\rangle\}_{A}\otimes\{|\rm{e}\rangle,|\rm{o}\rangle\}_{B}$, around 0-$\pi$ point($0.48<\phi_{eA(B)}<0.52$), the Hamiltonian takes the form
\begin{equation}
    H=\frac{1}{2}\delta E_{A}\sigma_z^1+\frac{1}{2}\delta E_{B}\sigma_z^2+E_{JA}^{(1)}(\phi_{eA})\sigma_x^1\sigma_0^2+E_{JB}^{(1)}(\phi_{eB})\sigma_0^1\sigma_x^2,
\label{eq4}
\end{equation}
where the superscripts (1,2) label the $0-\pi$ qubits, and the subscript(0,x,y,z) are the index of palui matrix. $\delta E_{A(B)}$ takes the form\cite{Smith2020}
\begin{equation}
    \delta E_{i}=16E_{ci}\sqrt{\frac{2}{\pi}}(\frac{2E_{J}^{(2)}(\phi_{ei})}{E_{ci}})^{\frac{3}{4}}e^{-\sqrt{\frac{2E_{J}^{(2)}(\phi_{ei})}{E_{ci}}}}\cos(\pi n_{gi}),
\label{eq5}
\end{equation}
here $i$ can be $A$ or $B$. For simplification, we write it as $\delta E_i=F(E_{ci},\phi_{ei})\cos(\pi n_{gi})$ with $F(E_{ci},\phi_{ei})=16E_{ci}\sqrt{\frac{2}{\pi}}(\frac{2E_{J}^{(2)}(\phi_{ei})}{E_{ci}})^{\frac{3}{4}}e^{-\sqrt{\frac{2E_{J}^{(2)}(\phi_{ei})}{E_{ci}}}}$. Firstly, we assume the circuit elements, occurring pairwise, are identical in the ideal case, say $E_{cA}=E_{cB}=E_c,n_{gA}=n_{gB}=n_g,\phi_{eA}=\phi_{eB}=\phi_e=0.5$. With the Hamiltonian (Eq.~(\ref{eq4})) and the expression of the coefficients (Eq.~(\ref{eq2}),(\ref{eq5})), we can get the eigenvalues of the four states change with $n_g$ at $\phi_{eA}=\phi_{eB}=0.5$(the solid lines in Fig.~\ref{fig-3}(a)), it has two flat energy levels ($|\rm{eo}\rangle,|\rm{oe}\rangle$) and can be encoded as qubit states. The corresponding wavefunctions are shown in Fig.~\ref{fig-3}(c). This indicates that the middle two states are completely immune to charge noise, similar to the S-T qubit being immune to global uniform magnetic fields\cite{Petta2005}. Apart from that,  compairing with Fig.~\ref{fig-2}(f), the middle two states change slightly with flux at half flux quantum (solid lines in Fig.~\ref{fig-3}(b)). Because when the flux $\phi_e$ deviate from $0.5\phi_0$, the Hamiltonian(Eq.~(\ref{eq3})) will get finite term $E_{J}^{(1)}(0.5+\delta\phi_e)\cos(\hat{\phi}_1)$ or $E_{J}^{(1)}(0.5+\delta\phi_e)\cos(\hat{\phi}_2)$. Though $\cos\hat{\phi}_1(\hat{\phi}_2)$ can couple the $|\rm{e}\rangle,|\rm{o}\rangle$ states in A(B) subspace, it can not couple the qubit states $|\rm{eo}\rangle,|\rm{oe}\rangle$ directly. Instead, they can only affect the qubit states via other energy levels in a second-order process rather than first order process, which can suppress the flux noise.

However, in practice, the circuit elements may not be exactly identical
. There may exist parameters differences in the circuit elements. Here, the consistency of Josephson coupling energy ($E_J^{(2)}$) is guaranteed by the 2DTI Josephson junctions as discussed in part II and we do not discuss here. For the capacitance difference ($C_{gA(B)}$), it will affect the charge energy $E_{cA(B)}$ and offset charge $n_{gA(B)}$ simultaneously. In practice, in order to achieve the weak transmon regime, the Josephson Junctions are often shunted with a large capacitor, $C_{gA(B)}\gg C_{JA(B)}$, to reduce the charge energy\cite{Krantz2019}. As a result, we ignore the capacitance difference of Josephson junction and focus on the capacitance difference of the shunted capacitor. Here we define $C_{gA(B)}=C_g\pm\delta C_g$, and $\delta C_g/C_g\equiv x$. The difference of capacitance will lead to the charge energy difference of the two 0-$\pi$ qubits, analogous to the difference in the g-factor in the S-T qubit scenario. The two states ($|\rm{eo}\rangle,|\rm{oe\rangle}$) will gets finite gap at $n_g=0$ and half flux quantum, as shown in Fig.~\ref{fig-3}(a)(b)(the dashed lines) with $x$ up to 5\%. But it is still a sweet spot for the offset charge $n_g$ and flux $\phi_{eA(B)}$, demonstrating the robustness of the proposed architecture.

To clarify it clearly, we project the Hamiltonian Eq.~\eqref{eq4} into the qubit space $\{|\rm{eo}\rangle,|\rm{oe}\rangle\}$ as 
\begin{equation}
    \tilde{H}=\frac{1}{2}(\delta E_A-\delta E_B)s_z+\frac{2[E_{JA}^{(1)}(\phi_{eA})^2-E_{JB}^{(1)}(\phi_{eB})^2]}{\delta E_A+\delta E_B}s_z,
\label{eq6}
\end{equation}
with $s$ the Pauli matrix acting on the qubit space. In the ideal case, $n_{gA}=n_{gB}=n_g, \delta E_A=\delta E_B$. It clearly shows that the qubit states are degenerate when the flux $\phi_{eA}=\phi_{eB}=0.5$. The energy of the qubit states are stable with the offset charge $n_g$(two flat lines in Fig.~\ref{fig-3}(a)). It can only change the energy difference between the two degenerate states and other states, similar to the effect of a uniform magnetic field in the S-T qubit. Therefore, in principle, we can get an infinite $T_2^{n_g}$ for the two degenerate states. Then, we consider the flux noise effect at $n_g=0$. Since the flux $\phi_{eA},\phi_{eB}$ can be tuned independently. We fix $\phi_{eA}=0.5$ and calculate the energy level changes with $\phi_{eB}$. Then the Hamiltonian (Eq.~(\ref{eq6})) becomes the form
\begin{equation}
        \tilde{H}=-\frac{2E_{JB}^{(1)}(\phi_{eB})^2}{\delta E_A+\delta E_B}s_z.
        \label{eq-phi}
\end{equation}
Consequently, the energy of the qubit states becomes parabolic and touches at $\phi_{eB}=0.5$(solid lines in Fig.~\ref{fig-3}(b)). It is because the flux noise can not directly couple the qubit states with the total parity odd, it can only couple the qubit states with other states. The coefficient $\frac{2E_{JB}^{(1)}(\phi_{eB})^2}{\delta E_A+\delta E_B}$ becomes parabolic change with flux at $\phi_{eB}=0.5$, rather than linear change due to second-order perturbation. Additionally, the first-order derivative of $E_J^{(1)}(\phi_{eB})^2$ is zero, not maximum at $\phi_e=0.5$. This property leads to a significant enhancement of $T_2^{(\phi_e)}$, as compared to the case shown in Fig.\ref{fig-2}(f). And the conclusion holds if we fix $\phi_{eB}=0.5$ change $\phi_{eA}$.

For the capacitance difference in the circuit elements, the offset charge takes the form $n_{gA(B)}=n_g(1\pm \delta C_g/C_g)$ with $n_g=C_gV/2e$, and the charge energy takes the form  $E_{cA(B)}=E_c/(1\pm\delta C_g/C_g)$. The effective Hamiltonian Eq.~(\ref{eq6}) becomes the form 
\begin{equation}
\begin{split}
    \tilde{H}=&\frac{1}{2}[F(E_c/(1+x),\phi_{eA})\cos(\pi n_g(1+x))\\
    &-F(E_c/(1-x),\phi_{eB})\cos(\pi n_g(1-x))]s_z\\
    &+\frac{2[E_{JA}^{(1)}(\phi_{eA})^2-E_{JB}^{(1)}(\phi_{eB})^2]}{\delta E_A+\delta E_B}s_z
    \label{eq-ng}.
\end{split}
\end{equation}
It clearly shows that the difference in capacitance can cause a finite gap between the two degenerate states, as shown in Fig.~\ref{fig-3}(a)(b)(the dashed lines), and the energy gap is about 0.2 GHz for $x$ up to 5\%. It is still sweet spots and maintains resistance
to charge noise and flux noise, simultaneously.

\subsection{Coherent properties of the qubit}
In this part, we calculate the coherent properties of the qubit. As the qubit is tuned by flux and offset charge, so we mainly consider the flux noise and charge noise. We thus expand the Hamiltonian up to the second order at the sweet spots
\begin{equation}
    H=H_0+\frac{\partial H}{\partial \lambda}\delta\lambda(t)+\frac{1}{2}\frac{\partial^2H}{\partial\lambda^2}\delta\lambda^2(t),
\end{equation}
where $H_0$ is the ideal Hamiltonian at the sweet spots, and $\lambda$ can be charge ($n_g$) or flux ($\phi_e$).

With the Hamiltonian, we can get transition rate with Fermi's golden rule\cite{Griffiths2018}, it takes the form 
\begin{equation}
    \Gamma_{i\to f} = |\langle \psi_f| \frac{\partial H}{\partial \lambda}|\psi_i\rangle|^2 S_{\lambda}(\omega_{fi}),
\end{equation}
where $\phi_{i(f)}$ is the initial(final) state, $S_{\lambda}(\omega)$ is the noise power spectrum.


\begin{figure}[!htbp]
	\centering
	\includegraphics[width=1\columnwidth]{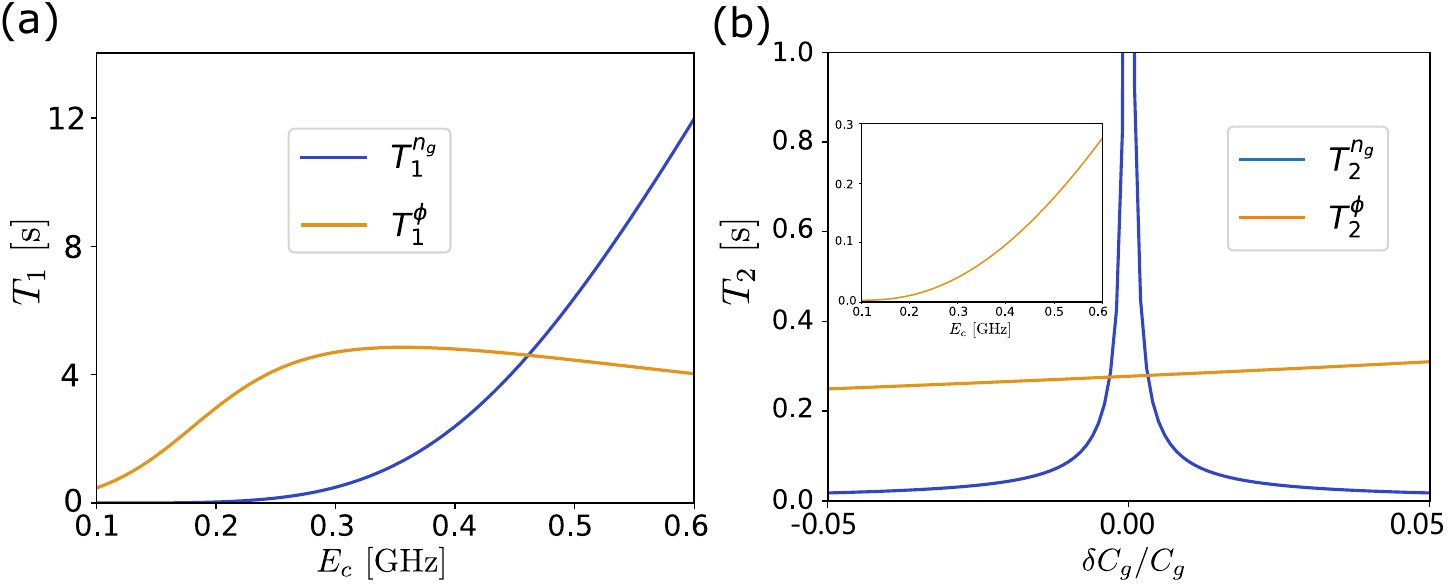}
	\caption{(a) Relaxation time changes with charge energy $E_c$. (b) Dephasing time changes with capacitance difference $\delta C_g/C_g$ with $E_c=0.6$ GHz. At $\delta C_g=0$, $T_2^{n_g}$ is infinity. Inset shows $T_2^{\phi}$ changes with charge energy $E_c$ at $\delta C_g=0$.}
\label{fig-4}
\end{figure}
Firstly, we consider the charge noise. Charge noise comes from the fluctuation of voltage which is capacitively coupled to the circuit(Fig.~\ref{fig-1}(a)). Though there are $n_{gA(B)}$ in the Hamiltonian, it is tuned by the same voltage $V$, similar to a uniform magnetic field in the S-T qubit. With the Hamiltonian Eq.~(\ref{eq3}), we can get
\begin{equation}
    \frac{\partial H}{\partial V}=8E_{cA}\frac{C_{gA}}{2e}\hat{n}_{1}+8E_{cB}\frac{C_{gB}}{2e}\hat{n}_{2}=2e(\hat{n}_{1}+\hat{n}_{2}),
\end{equation}
where $E_{cA(B)}C_{gA(B)}=e^2/2$ in the limit $C_{gA(B)}\gg C_{JA(B)}$. As the operator $\hat{n}_{1(2)}$ can not change the parity of the states in each subspace, it can only cause the transition of states with the same parity in each subspace. So, for ${|\rm{eo}\rangle,|\rm{oe}\rangle}$ qubit states, it can only cause the transitions of the qubit states to higher levels. The transition rate takes the form\cite{Groszkowski2018}
\begin{equation}
 \Gamma_{1}^{n_g}=\Gamma_{|\rm{eo}\rangle\to|\rm{A_3o}\rangle}+\Gamma_{|\rm{oe}\rangle\to|\rm{A_2e}\rangle}+\Gamma_{|\rm{eo}\rangle\to|\rm{eB_2}\rangle}+\Gamma_{|\rm{oe}\rangle\to|\rm{oB_3}\rangle},
\end{equation}
where $\rm{A(B)_{2(3)}}$ represent the second(third) excite state in subspace A or B. The spectrum in the cavity takes the form\cite{Smith2020}
\begin{equation}
    S(\omega)=\frac{2}{C_JQ_{cap}(\omega)}P(\omega)
\end{equation}
where $C_J=e^2/2E_C$ is the capacitance, $P(\omega)$ is the statistic distribution, and it takes the form $1/(e^{\hbar\omega/k_BT}-1)$ for the "up transition" and $e^{\hbar\omega/k_BT}/(e^{\hbar\omega/k_BT}-1)$ for the "down transition"\cite{Smith2020, Krantz2019}. Here, $\rm{T}$ is the temperature, $Q_{cap}(\omega)=Q_{cap}(\frac{6~\mathrm{GHz}}{|\omega|})^{0.7}$, the nominal value $Q_{cap}\sim 1\times 10^6$ correspond to the resonant frequency of 6 GHz\cite{Smith2020}.

Next, we consider the effect of flux noise. In our device, we assume that the two fluxes ($\phi_{eA(B)}$) can be independently tuned, so the flux noise can be expressed in the form
\begin{equation}
    \frac{\partial H}{\partial\phi_{eA(B)}}=\frac{\partial E_{JA(B)}^{(1)}}{\partial\phi_{eA(B)}}\cos\hat{\phi}_1(\hat{\phi}_2),
\end{equation}
which can flip the parity of the state in each subspace, causing the transition between qubit states and other energy levels. The transition takes the form\cite{Groszkowski2018}
\begin{equation}
    \Gamma_{\phi}^{1}=\Gamma_{|\rm{eo}\rangle\to|\rm{ee}\rangle}+\Gamma_{|\rm{oe}\rangle\to|\rm{oo}\rangle}.
\end{equation}
For flux noise, we take the noise spectrum as the form $S_{\phi_e}(\omega)=2\pi \frac{A_{\phi_e}}{|\omega|}P(\omega)$ with $(\omega_{ir}<\omega<\omega_{uv})$\cite{Ithier2005, Paladino2014, Krantz2019}. $A_{\phi_e}$ is related to the circumference of the device \cite{Faoro2008, Lanting2009}, for the device with 2DTI, the circumference can be two orders of magnitude less than traditional SQUID device\cite{Blais2007, Roessler2023}. Therefore, we take $A_{\phi_e}=10^{-14} \phi_0^2$ for flux noise. Then we can get the $T_1$ of the qubit, as shown in Fig.~\ref{fig-4}(a).

The dephasing time $T_2$ is related to the decay of the off-diagonal term of the density matrix\cite{Koch2007, Krantz2019},
\begin{equation}
    \rho_{01}=\exp(-iD_1\int_0^t \delta \lambda(t)dt-i \frac{1}{2}D_2\int_0^t\delta \lambda(t)^2dt),
\end{equation}
where $D_1 = \frac{\partial \omega}{\partial \lambda}$ and $D_2 =\frac{\partial^2 \omega}{\partial \lambda^2}$. And for $1/f$ noise $S_{\lambda}(\omega)=2\pi A_\lambda/|\omega|(\omega_{ir}<\omega<\omega_{uv})$\cite{Ithier2005,Paladino2014}, $\lambda$ can represent charge or flux. Here we estimate that $\omega_{ir}/2\pi=1$ Hz, $\omega_{uv}/2\pi=0.4$ GHz which is determined by temperature of the system ($T_m<20$ mK), $A_{n_g}=10^{-8}$ $e^2$ \cite{Koch2007}. The dephasing time takes the form
\begin{equation}
    T_2 = [D_2^2A_\lambda^2 \ln^2(\frac{\omega_{uv}}{\omega_{ir}})+2D_2^2A_\lambda^2\ln^2(\frac{1}{\omega_{ir}t})]^{-\frac{1}{2}}. 
\label{t2}
\end{equation}
Notice that the energy of states $|\rm{eo}\rangle,|\rm{oe}\rangle$ are independent of $n_g$ if the capacitances $C_{gA(B)}$ are the same, and we expect infinite $T_2^{n_g}$ in principle. However, with capacitance difference, we can get finite $T_2^{n_g}$ due to the voltage fluctuations. With Eq.~\ref{eq-ng}, we can get the expression of $\partial^2\omega/\partial V^2$ at sweet spot of voltage
\begin{equation}
    \frac{\partial^2\omega}{\partial(\frac{C_gV}{2e})^2}= -(1+\frac{\delta C_g}{C_g})^2\pi^2F(E_{cA})+(1-\frac{\delta C_g}{C_g})^2\pi^2F(E_{cB}),
\end{equation}
the capacitance difference $\delta C_g/C_g\ll1$, and we use the $n_g=C_gV/2e$ to calculate the dephasing time $T_2^{n_g}$, the result is shown in Fig.~\ref{fig-4}(b). For $T_2^{\phi}$, the working point is a sweet spot for $\phi_{eA(B)}$. Moreover, with Eq.~(\ref{eq-phi}), we can get the conclusion that the second-order derivative $\partial^2\omega/\partial\phi^2$ decreases with increasing charge energy $E_c$, leading to larger $T_2^{\phi_e}$. This effect is similar to increasing the Zeeman field by increasing the g-factor in the S-T qubit. The result is shown in the inset of Fig.~\ref{fig-4}(b). Overall, we estimate the relaxation time $T_1=4.6$ s, and the dephasing time $T_2=147$ ms at $E_c=0.45$ GHz.

\subsection{Operation of the qubit}
With the qubit states, we then consider the operation of the qubit. We begin with the effective Hamiltonian (Eq.~(\ref{eq6})). In the ideal case, $\phi_{eA}=\phi_{eB}=\phi_e=\frac{1}{2}\phi_0$, $\delta E_A=\delta E_B=\delta E$, and thus $\tilde{H}=0$. In experiment, We can slightly tune the flux($\phi_{eA(B)}$) slightly away from half flux quantum and it will generate $\cos\hat{\phi}_1(\hat{\phi}_2)$ term in the Hamiltonian. These terms can couple the qubit states with other energy levels. By employing Eq.~(\ref{eq2}) and Eq.~(\ref{eq6}), the effective Hamiltonian of the qubit becomes 
\begin{figure}[!htbp]
	\centering
	\includegraphics[width=1\columnwidth]{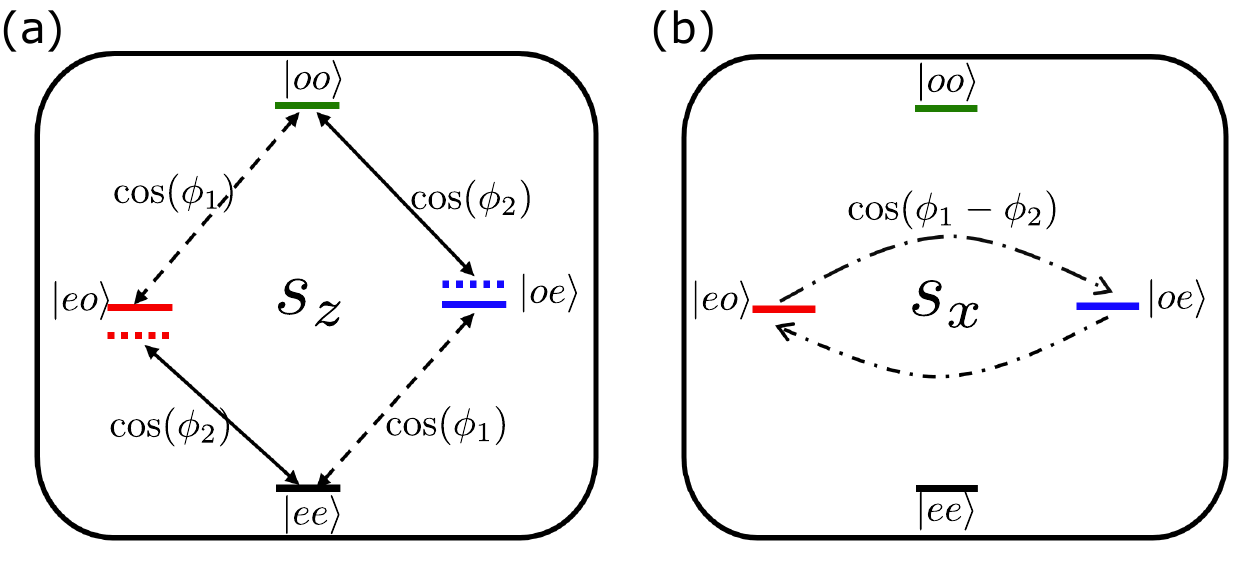}
	\caption{(a) Schematic diagram of the coupling term $\cos(\phi_1)$ and $\cos(\phi_2)$, it can afford rotations around the z-axis in qubit space. (b) Schematic diagram of the coupling term $\cos(\phi_1-\phi_2)$, it can afford rotations around the x-axis in qubit space.}
\label{fig-5}
\end{figure}

\begin{equation}
\begin{aligned}
    \tilde{H_z}&=\pm\frac{625[\cos(2\pi\phi_e)-\cos(2\pi(\phi_e+\delta\phi_e))]}{2\delta E}s_z\\
    &\approx\pm\frac{1250\pi\delta\phi_e}{2\delta E}s_z=h_z(\delta\phi_e)s_z,
\label{eq7}
\end{aligned}
\end{equation}
where $\pm$ represents the local tuning of the flux $\phi_{eA}$ or $\phi_{eB}$. Consequently, the time evolution of the effective Hamiltonian results in Z rotations of the qubit states (Fig.~\ref{fig-5}(a)). Additionally, we can close the switch(tune the gate voltage of the gatemon) and connect the circuit node with a Josephson Junction described by $E_J\cos(\hat{\phi}_1-\hat{\phi}_2)$ (Fig.~\ref{fig-5}(b)), which will flip the parity of the two 0-$\pi$ qubits, simultaneously. Here we ignore the small capacitance of the connecting Junction\cite{supp}.
In basis $\{|\rm{e}\rangle,|\rm{o}\rangle\}\otimes\{|\rm{e}\rangle,|\rm{o}\rangle\}$, it takes the form $E_J\sigma_x^1\sigma_x^2$. Similarly, we can do second-order perturbation theory to project the coupling Hamiltonian into qubit subspace $\{|\rm{eo}\rangle,|\rm{oe}\rangle\}$, and the effective coupling Hamiltonian takes the form
\begin{equation}
    \tilde{H_x}=h_x(E_J)s_x,
\label{eq8}
\end{equation}
where $h_x(E_J)=E_J$, the time evolution of the coupling Hamiltonian will be an X rotation for the qubit states (Fig.~\ref{fig-4}(b)), achieving electric control of the qubit. Combining Eq.~(\ref{eq7}) and Eq.~(\ref{eq8}), we obtain the qubit operation Hamiltonian
\begin{equation}
H_{\rm{control}}=h_z(\delta\phi_e)s_z+h_x(E_J)s_x.
\end{equation}
With the operation Hamiltonian, we can implement arbitrary gates for the qubit. For qubit initialization, we can begin with the lowest energy state $|\rm{ee}\rangle$ of the system, then slightly tune the flux in one 2DTI Josephson Junction to introduce the term $E_{J}^{(1)}(\delta \phi_e)\cos(\hat{\phi}_1)$, $\delta\phi_e$ is the flux
offsets from half-flux quantum. It can couple the states $|\rm{ee}\rangle,|\rm{oe}\rangle$, with $\eta=\langle \rm{ee}|E_{J}^{(1)}(\delta\phi_e)\cos(\hat{\phi}_1)|\rm{oe}\rangle$. After free time evolution for a duration $t=\pi/2\eta$, we can gain the initial state $|\rm{oe}\rangle$ state. The initial state $|\rm{eo}\rangle$ can be gained by shifting the flux of the other TI Junction with the same process. The readout scheme for the parity-protected qubit has been developed and can be employed in our proposal\cite{Larsen2020, Gyenis2021,Schrade2022}.

We then do a discussion on the experiment of our proposal. (1) Edge length difference in one TI Josephson Junction. The junction length of the two effective Josephson Junctions contributed by the helical edge states may not be the same. Then, at half flux quantum, there may exist finite $\sin(\hat{\phi})$ term for each 2DTI Josephson Junction. However, as the device is worked at $n_g=0$, finite $\sin(\hat{\phi})$ term does not affect parity protected states\cite{Maiani2022,Guo2022}, the influence can be safely neglected. (2) Junction length difference between two 2DTI Josephson Junctions. It can cause Josephson coupling ($E_J^{(2)}$) difference between them. Assuming the length difference is about $1\%\xi$, then with Fig.~\ref{fig-2}(d) and Eq.~(\ref{eq5}), we can get the gap of the qubit states induced by the Junction length difference is the order of $10^{-2}$ GHz. This gap is significantly small and can not change the degeneracy of the qubit states. (3) Bulk states in 2DTI Josephson Junction. The presence of bulk states in the 2DTI Josephson Junction can introduce finite $\cos(\hat{\phi})$ term for each Junction. However, as shown in Eq.~(\ref{eq6}), the $\cos(\hat{\phi})$ term of the two Junctions are canceled with each other, and they do not affect the qubit states. Moreover, with the recent experiment, the topological insulator can be gained with the material  $\mathrm{Bi}_{4}\mathrm{Br}_{4}$ which has a large bulk gap (0.2 eV) and clean bulk states\cite{Yang2022, Shumiya2022}. This is favorable for the qubit design.

In conclusion, we have proposed a novel architecture to implement a parity-protected qubit with two 0-$\pi$ qubits, which exhibits similarities to the singlet-triplet qubit. The qubit states are immune to offset charge in the ideal case which makes it robust with offset charge $n_g$. Moreover, the qubit states change slowly with flux compared to a single 0-$\pi$ qubit, resulting in an increased dephasing time $T_2^{\phi_e}$. By leveraging the consistency of TI Josephson Junctions, we can implement this device experimentally, providing a reliable and promising platform for achieving parity-protected qubit states.

%

\end{document}